\documentclass[sigconf, authorversion]{acmart} 

\usepackage[frozencache,cachedir=.]{minted}
\usepackage{xcolor}
\usepackage{multicol}
\usepackage{enumitem}
\usepackage{todonotes}
\usepackage{rotating}
\usepackage{multirow}
\usepackage{natbib}
\usepackage{wasysym}
\RequirePackage[skins]{tcolorbox}
\newtcolorbox{custombox}[1]{
	colback=white,
	colframe=white,
	left=1mm,
	right=1mm,
	top=1mm,
	bottom=1mm,
	fonttitle=\bfseries,
	arc=0mm,
	leftrule=1mm,
	rightrule=0mm,
	toprule=0mm,
	bottomrule=0mm,
	notitle,
	before=\par\smallskip\noindent,
	before upper={\textbf{#1: } },
}

\AtBeginDocument{%
  \providecommand\BibTeX{{%
    \normalfont B\kern-0.5em{\scshape i\kern-0.25em b}\kern-0.8em\TeX}}}

\copyrightyear{2024}
\acmYear{2024}
\setcopyright{rightsretained}
\acmConference[SIGCSE 2024]{Proceedings of the 55th ACM Technical Symposium on Computer Science Education V. 1}{March 20--23, 2024}{Portland, OR, USA}
\acmBooktitle{Proceedings of the 55th ACM Technical Symposium on Computer Science Education V. 1 (SIGCSE 2024), March 20--23, 2024, Portland, OR, USA}
\acmDOI{10.1145/3626252.3630909}
\acmISBN{979-8-4007-0423-9/24/03}

\begin{document}
\pagestyle{plain} 



\title{Prompt Problems: A New Programming Exercise for the Generative AI Era}


\author{Paul Denny}
\orcid{0000-0002-5150-9806}
\affiliation{
  \institution{The University of Auckland}
  \city{Auckland}
  \country{New Zealand}
}
\email{paul@cs.auckland.ac.nz}

\author{Juho Leinonen}
\orcid{0000-0001-6829-9449}
\affiliation{
  \institution{The University of Auckland}
  \city{Auckland}
  \country{New Zealand}
}
\email{juho.leinonen@auckland.ac.nz}

\author{James Prather}
\orcid{0000-0003-2807-6042}
\affiliation{
  \institution{Abilene Christian University}
  \city{Abilene, TX}
  \country{USA}
}
\email{james.prather@acu.edu}

\author{Andrew Luxton-Reilly}
\orcid{0000-0001-8269-2909}
\affiliation{
  \institution{The University of Auckland}
  \city{Auckland}
  \country{New Zealand}
}
\email{a.luxton-reilly@auckland.ac.nz}

\author{Thezyrie Amarouche}
\orcid{0000-0003-3725-0049}
\affiliation{
  \institution{University of Toronto Scarborough}
  \city{Toronto, ON}
  \country{Canada}
}
\email{thezyrie.amarouche@mail.utoronto.ca}

\author{Brett A. Becker}
\orcid{0000-0003-1446-647X}
\affiliation{
  \institution{University College Dublin}
  \city{Dublin}
  \country{Ireland}
}
\email{brett.becker@ucd.ie}

\author{Brent N. Reeves}
\orcid{0000-0001-5781-1136}
\affiliation{%
  \institution{Abilene Christian University}
  \city{Abilene, Texas}
  \country{USA}
}
\email{brent.reeves@acu.edu}

\renewcommand{\shortauthors}{Denny, Leinonen, Prather, et al.}

\begin{abstract}

Large Language Models (LLMs) are revolutionizing the field of computing education with their powerful code-generating capabilities. Traditional pedagogical practices have focused on code \emph{writing} tasks, but there is now a shift in importance towards code reading, comprehension and evaluation of LLM-generated code. Alongside this shift, an important new skill is emerging -- the ability to solve programming tasks by constructing good prompts for code-generating models. In this work we introduce a new type of programming exercise to hone this nascent skill: `Prompt Problems'. Prompt Problems are designed to help students learn how to write effective prompts for AI code generators. A student solves a Prompt Problem by crafting a natural language prompt which, when provided as input to an LLM, outputs code that successfully solves a specified programming task. We also present a new web-based tool called \textsc{Promptly} which hosts a repository of Prompt Problems and supports the automated evaluation of prompt-generated code. We deploy \textsc{Promptly} for the first time in one CS1 and one CS2 course and describe our experiences, which include student perceptions of this new type of activity and their interactions with the tool. We find that students are enthusiastic about Prompt Problems, and appreciate how the problems engage their computational thinking skills and expose them to new programming constructs. We discuss ideas for the future development of new variations of Prompt Problems, and the need to carefully study their integration into classroom practice.

\end{abstract}




\maketitle

%
%
\section{Introduction}


The advent of large language models (LLMs) that can generate code is having a rapid and significant impact on computing education practice, particularly at the introductory level \cite{prather2023robots}.  Traditional pedagogical approaches have focused on helping students learn how to \emph{write} code.  This is typically achieved through frequent practice involving many small problems \cite{allen2019many, denny2011codewrite} or through scaffolding via activities such as Parsons problems \cite{ericson2022parsons, du2020review}.  However, LLMs are now capable of producing code automatically and have demonstrated impressive performance on problems that are typical in introductory programming courses \cite{finnie-ansley2022robots, finnie-ansley2023my, reeves2023evaluating}. 
In addition to the opportunities they present, educators have voiced concerns around the potential misuse of these models for plagiarism, and over-reliance on AI-generated code by beginners~\cite{becker2023programming}, leading to a possible erosion of traditional coding skills \cite{denny2023computing}. New pedagogical approaches are needed to develop the changing skillsets that students require in the era of generative AI \cite{denny2023chat}. 


Teaching students to read and understand code are longstanding goals of introductory courses, and they are becoming increasingly important skills given the ease with which code can be generated by LLM-based tools. 
An equally important emerging skill is \emph{the ability to formulate effective prompts for LLMs to generate code}. Recent work has shown that although many typical introductory problems can be solved by LLMs using verbatim textbook or exam problem statements~\cite{finnie-ansley2022robots, finnie-ansley2023my}, this approach is not always sufficient.  For example, manual modification of the prompts to include explicit algorithmic hints greatly improves code-generation performance \cite{tang2022solving}.  In recent work, Denny et al. argue that the ability to engineer effective prompts that generate correct solutions is now an essential skill for students, yet they do not propose concrete approaches for how this skill can be taught \cite{denny2023conversing}.

To address this concern, we introduce the concept of a `Prompt Problem' -- a new exercise paradigm in which students solve programming exercises by formulating natural language prompts for code-generating LLMs.  Students are presented with a representation of a problem that illustrates how input values should be transformed to an output.  Their task is to devise a prompt that would guide an LLM to generate the code required to solve the problem.  


In addition to conceptualizing the problem type, we make two other contributions in this work: (1) we introduce a tool (called \textsc{Promptly}) for delivering Prompt Problems, that displays a problem representation, converts a prompt written by a student to code (via an API call to an LLM), and then executes the code against a suite of test cases; and (2) we present our observations from deploying Prompt Problems to programming students in a CS1 course and a CS2 course, and reflect on our experiences of using them in our teaching for the first time.

\section{Related work}
It has been less than a year since LLMs began to dominate conversations in the computing education community and a little more than that since the first research papers began to emerge in the computing education literature. Early work centered on the capabilities of these tools, largely driven by concerns that they would lead to a flood of cheating~\cite{malinka2023on} and the effect that would have on student learning. Sometimes, such work involved comparing LLM and student performance, for example in generating explanations of code~\cite{leinonen2023comparing}. Finnie-Ansley et al. demonstrated that Codex (based on GPT-3) ranked in the top quartile of real introductory programming (CS1) students on real exams~\cite{finnie-ansley2022robots}. A year later Finnie-Ansley et al. extended this work to data structures and algorithms (CS2) exams with very similar results~\cite{finnie-ansley2023my}. Other studies on the capabilities of LLMs have revealed impressive proficiency in dealing with object-oriented programming tasks~\cite{cipriano2023gpt-3}, Parsons problems~\cite{reeves2023evaluating}, and compiler error messages~\cite{leinonen2023using}. Many of these explorations also revealed that LLMs are not infallible and can produce solutions that do not align with best programming practice~\cite{cipriano2023gpt-3}, struggle with longer and higher-level specifications~\cite{finnie-ansley2022robots}, 
and cause students to become confused reading code that they did not write themselves~\cite{kazemitabaar2023studying, prather2023its}. Babe et al. showed that LLMs can mislead students, causing them to believe that their own prompts are more (or less) effective than they are in reality~\cite{babe2023studenteval}.

Recently, the focus has started to shift from assessing the capabilities of LLMs to using them in teaching and learning practice~\cite{macneil2023implications}. For example, Sarsa et al. showed that LLMs can generate viable programming exercises including test cases and explanations~\cite{sarsa2022automatic}, and Liffiton et al. describe the use of an LLM-powered teaching assistant with guardrails suitable for computing courses \cite{liffiton2023codehelp}. 
There has also been a recent focus on prompting strategies and understanding the prompts that students create.  
White et al. present a prompt structuring framework for constructing prompts that can be applied across problem domains, and demonstrate how prompts can be constructed from patterns \cite{white2023prompt}. 
A benchmark dataset of 1,749 prompts across 48 problems, written by 80 novice Python programming students~\cite{babe2023studenteval}, has recently been published which can be used by others for LLM benchmarking as well as tool development.

A logical next step towards integrating LLMs into teaching practice is developing tools and resources to aid students in effectively working with LLMs for learning. Lao and Guo interviewed 19 introductory programming instructors from nine countries across six continents and found that some instructors are embracing the idea of integrating AI tools into current courses via mechanisms such as giving personalized help to students and aiding instructors with time-consuming tasks~\cite{lau2023ban}. MacNeil et al. used LLM-generated code explanations successfully in a web software development e-book~\cite{macneil2023experiences}, and Zingaro and Porter are completing a textbook for teaching introductory programming with Copilot and ChatGPT from day one~\cite{Porter2023ai}. Integrating LLMs into computer science courses seems inevitable and stands to transform the way the subject is taught at all levels \cite{tedre2023k12computing,denny2023chat}. We believe that Prompt Problems will be one important step along the journey towards integrating the use of LLMs in computer science education.

\begin{figure}
\centering
  \includegraphics[width=.8\linewidth]{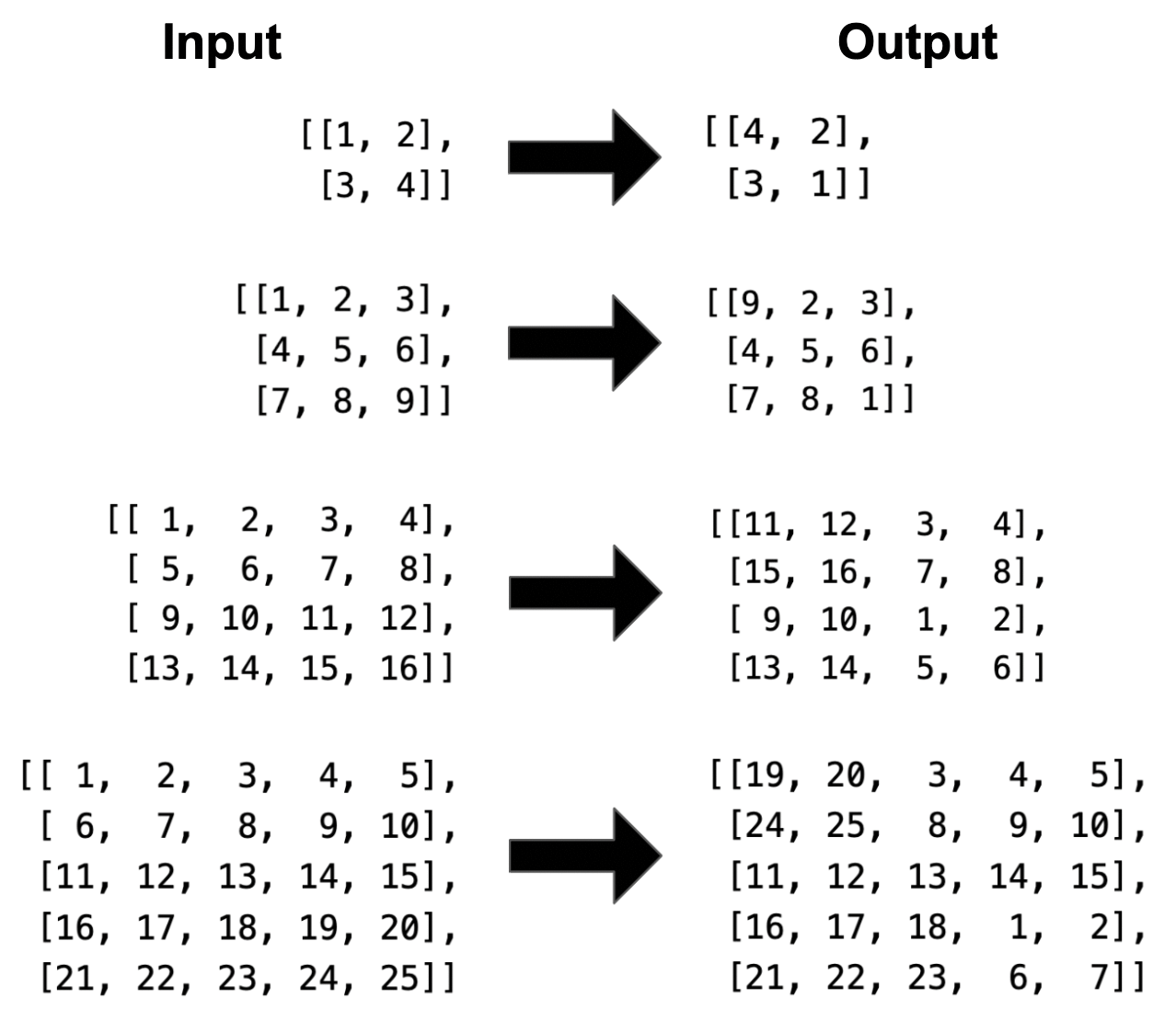}
  \caption{An example Prompt Problem that displays the data visually to prevent copying and pasting of the description into an LLM.  The goal is to swap the top-left and bottom-right non-overlapping quadrants of the matrix.}
  \label{fig:array-swap-problem}
\end{figure}

%
%
\section{Pilot Study}
\label{sec:pilot_study}

To motivate the need for our work, and to understand how students might try to use LLM tools like ChatGPT to communicate program requirements, we asked a group of graduate students the University of Auckland, New Zealand, to participate in a prompt writing assignment pilot study. This assignment took place during a single class session in April 2023. We provided a visual representation of a problem (see Figure \ref{fig:array-swap-problem}) and asked participants to query ChatGPT to write a program that could convert the shown inputs to the corresponding example outputs. The problem description was provided visually to prevent participants from easily copying and pasting it and, instead, to encourage them to formulate a suitable prompt themselves. 

Fifteen graduate students participated in the pilot, completing the activity described above, reflecting on it by writing an open-response review of the task, and opting to share their work with us. We expected computer science graduate students to have very few problems writing effective prompts, however this was not the case. Students wrote incomplete prompts (e.g. \emph{``I have a square matrix, and I want to swap the first half of the rows with the second half of the rows''}), tried to engage in conversations with the tool to refine the generated code, and tried to solve the wrong problem (e.g. \emph{``give me a function which works by first swapping the elements of each row in place, and then swapping the elements of each column in place''}). It became apparent to us that students, even at the graduate level, could benefit from explicit prompt writing practice that could teach them to understand the problem, write a single thorough prompt, and check the code generated by the LLM as having complete test case coverage. We therefore propose the idea of Prompt Problems to address this new gap in programming education.

%
%
\section{Practicing Prompt Problems}
To support the delivery of Prompt Problems to students, we developed a web-based tool called \textsc{Promptly}.  Currently, this tool supports only one type of Prompt Problem, in which the code generated by the LLM is not editable by the learner so the prompt must be complete and self-contained.
Other variations are certainly possible and we discuss these in Section \ref{discuss:variations}.

\begin{figure}
\centering
  \includegraphics[width=.9\linewidth]{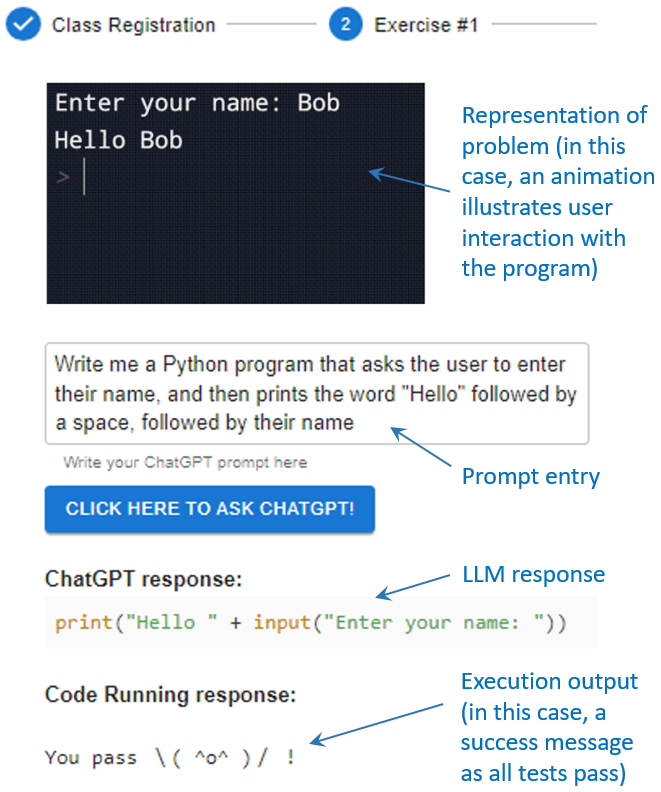}
  \caption{Interface layout for a Prompt Problem within the web-based \textsc{Promptly} tool (with figure annotations added in blue).  The layout is compressed for space reasons.}
  \label{fig:ex1_screenshot}
\end{figure}


\subsection{Tool Design}

Within the \textsc{Promptly} tool, sets of Prompt Problems are organized into course repositories which students select after logging in.  Each Prompt Problem within a course repository consists of a visual representation of a problem -- an image that does not include a textual description of the problem -- and a set of associated test cases that are used to verify the code that is generated by the LLM.  

When viewing a Prompt Problem, the student is shown the visual representation of the problem, and a partial prompt to complete.  For problems where the solution is a Python program, this partial prompt begins: ``Write a Python program that...'', to guide the student.  If the problem requires students to write a single function, then the partial prompt is: ``Write a Python function called...''.  When any text is entered, the ``Click here to ask ChatGPT!'' button is enabled, and clicking this button constructs a prompt that is sent to the LLM. This prompt consists of the verbatim text entered by the student, as well as some additional prompting to guide the model to produce only code and no additional explanatory text.  




Once the response is received from the LLM, it is then sent to a sandbox for execution against a test suite.  We use the publicly available sandbox associated with the CodeRunner tool \cite{lobb2016Coderunner}.  If the generated code passes the tests for the prompt problem, then the student receives a success message and is directed to progress to the next problem.  If any of the test cases fail, then the first failing test case is shown to the student.  They are then able to edit the prompt and resubmit in order to generate a new code response. 

Figure \ref{fig:ex1_screenshot} shows a screenshot of the tool interface (slightly compressed for space reasons).  In the screenshot, the learner has logged in, selected their course and exercise, and has entered a prompt that successfully solves the problem.  In our implementation, students must solve each problem in order to progress to the next problem. 

\begin{table*}[htb]
\small
\caption{Summary of student interactions with the Prompt Problems.  For each problem, a brief description and example is shown (the description is for the benefit of the reader and was not presented to students).  The total number of students (Students) who successfully solved each problem is given (the \% shown in parentheses is the percentage of students attempting the problem who successfully solved it).  Also shown is the average number of submissions (Sub) these students required, as well as the mean, minimum and maximum number of words used in successful prompts.}
\begin{tabular}{cllccccc}
{\bf Problem} & {\bf Description} & {\bf Example} & {\bf Students} & {\bf Sub}  & {\bf Mean} & {\bf Min} & {\bf Max}\\
\toprule
CS1-1 & Display a greeting and the user's name & Input: Serena $\rightarrow$ Hello Serena  & 44 (76\%) & 2.3 & 18.0 & 7 & 33 \\
CS1-2  & Classify an age using a set of four labels &  Input: 14 $\rightarrow$ Teenager & 31 (86\%)  & 1.8 & 47.9 & 26 & 85 \\
CS1-3  & Average the 3 middle values in a set of 5 values &  Input: 8.0 9.5 7.5 6.0 9.0 $\rightarrow$ 8.17  & 20 (65\%) & 7.5 & 40.7 & 25 & 66 \\
\hline
CS2-1  & Count the number of occurrences of 0 in a list & counter([0, 2, 3, 4, 0]) $\rightarrow$ 2  & 136 (75\%) & 2.4 & 23.0 & 10 & 84 \\
CS2-2  & Extract the first letter of each word in input string & initials(`abc def ghi') $\rightarrow$ `ADG'  & 121 (96\%) & 1.3 & 28.3 & 12 & 88 \\
CS2-3  & Create a list with element occurrences equaling values  & repeat([2, 0, 1, 3]) $\rightarrow$ [2, 2, 1, 3, 3, 3]  & 114 (99\%) & 1.5 & 34.2 & 16 & 92 \\
\bottomrule
\label{tab:summary}
\end{tabular}
\end{table*}




\subsection{Classroom Evaluation}
%
%

Prompt Problems are a novel task for learners in programming courses, and we are interested in understanding what students think about them.
\emph{They are also novel for instructors} -- and so we are particularly interested in understanding whether the problems we have created are appropriately challenging.



We deployed \textsc{Promptly} as an ungraded (i.e. optional) laboratory task in two Python-based courses (one CS1 and one CS2) taught at the University of Auckland, New Zealand.  
The CS1 lab was conducted in the second week of the course, at which point students were writing single-file scripts, without the use of functions, and had learned about standard input and output, arithmetic, and conditional statements.  For the CS2 course, the lab was also conducted in the second week of the course and all students in this course were familiar with the concept of functions.

Three problems were available on \textsc{Promptly} for each course.  Table \ref{tab:summary} provides a very brief description of each problem (note, these descriptions were not shown to students but are listed here for the benefit of the reader) alongside one example that illustrates one input with a corresponding output.  The CS1 problems all required the generation of a program that processed standard input and printed output, whereas the CS2 problems all required a function that returned a value.  The first problem in the CS1 course was the problem previously illustrated in Figure \ref{fig:ex1_screenshot}.
To evaluate the first use of Prompt Problems in our teaching, we explore the following two questions around how students interact with the problems and their opinions on this new type of learning activity:


\begin{itemize}
    \item When solving Prompt Problems, how many attempts do students require and to what extent do successful prompts vary in terms of length?
    \item What are students' perceptions of Prompt Problems and on learning programming through constructing prompts for LLMs?
\end{itemize}

For the three Prompt Problems in each course we investigate the number of prompt submissions required to solve each one and the number of words used in the submitted prompts.  To gauge student perceptions of solving Prompt Problems,
students in both courses were invited to provide feedback on their experience.  This feedback was not graded, and was given in response to the following prompt: \emph{``We would appreciate hearing about your experiences completing the exercises and in particular, how you think the experience of writing prompts may help you to learn programming''}. 





\section{Experiences}



The courses in which Prompt Problems were used were taught in July 2023, and participation by students was optional.  A total of 58 (out of 414 enrolled) students in the CS1 course and 182 (out of 444 enrolled) students in the CS2 course chose to attempt at least one problem on \textsc{Promptly}.

\subsection{Student Interactions with Prompt Problems}



As summarized in Table \ref{tab:summary}, in the CS1 course participants submitted 2.3 attempts (on average) for Problem 1, 1.8 for Problem 2, and 7.5 for Problem 3.  Given that only students who were successful on Problems 1 and 2 progressed to Problem 3, this last problem appeared to be the most difficult.  
The visual representation of this problem showed a row of five people (stylized as judges of a competition) holding up score cards with the maximum and minimum scores crossed out. 
Listing \ref{listing:prompts_students} shows three prompts that were submitted by different students attempting Problem 3 in the CS1 course (CS1-3).  Some students found it difficult to infer the goal from the problem representation.  For example, in the first prompt shown in Listing \ref{listing:prompts_students} the student has incorrectly inferred that values included in the average calculation should be sufficiently close to their predecessors.  The length of this incorrect prompt is 101 words -- in comparison the lengths of the \emph{correct} prompts for this problem ranged from 25 to 66 words.


In the second example in Listing \ref{listing:prompts_students}, the student has not attempted to provide a prompt that demonstrates they have understood what the problem is asking, but instead they have created a prompt that simply parrots back to the tool the three example tests cases shown in the problem description.  The student then asks the model: \emph{``Can you please replicate this program?''}.  The student submitted this prompt four times in a row, but all attempts were unsuccessful.
Finally, the third example in Listing \ref{listing:prompts_students} is the shortest successful prompt that was submitted for this problem (25 words).

Overall, the average number of words in successful prompts for the three CS1 problems was 18.0, 47.9, and 40.7.  In comparison, average successful prompt lengths for the CS2 problems were 23.0, 28.3 and 34.2.  We observed a consistent reduction in the number of students solving subsequent problems in each course -- this was not unexpected given the optional nature of the activity.  Success rates were particularly high in the CS2 course, with almost all students who progressed to Problems 2 and 3 solving them (with, on average, fewer than two submissions).



\begin{figure}
\centering
  \includegraphics[width=\linewidth]{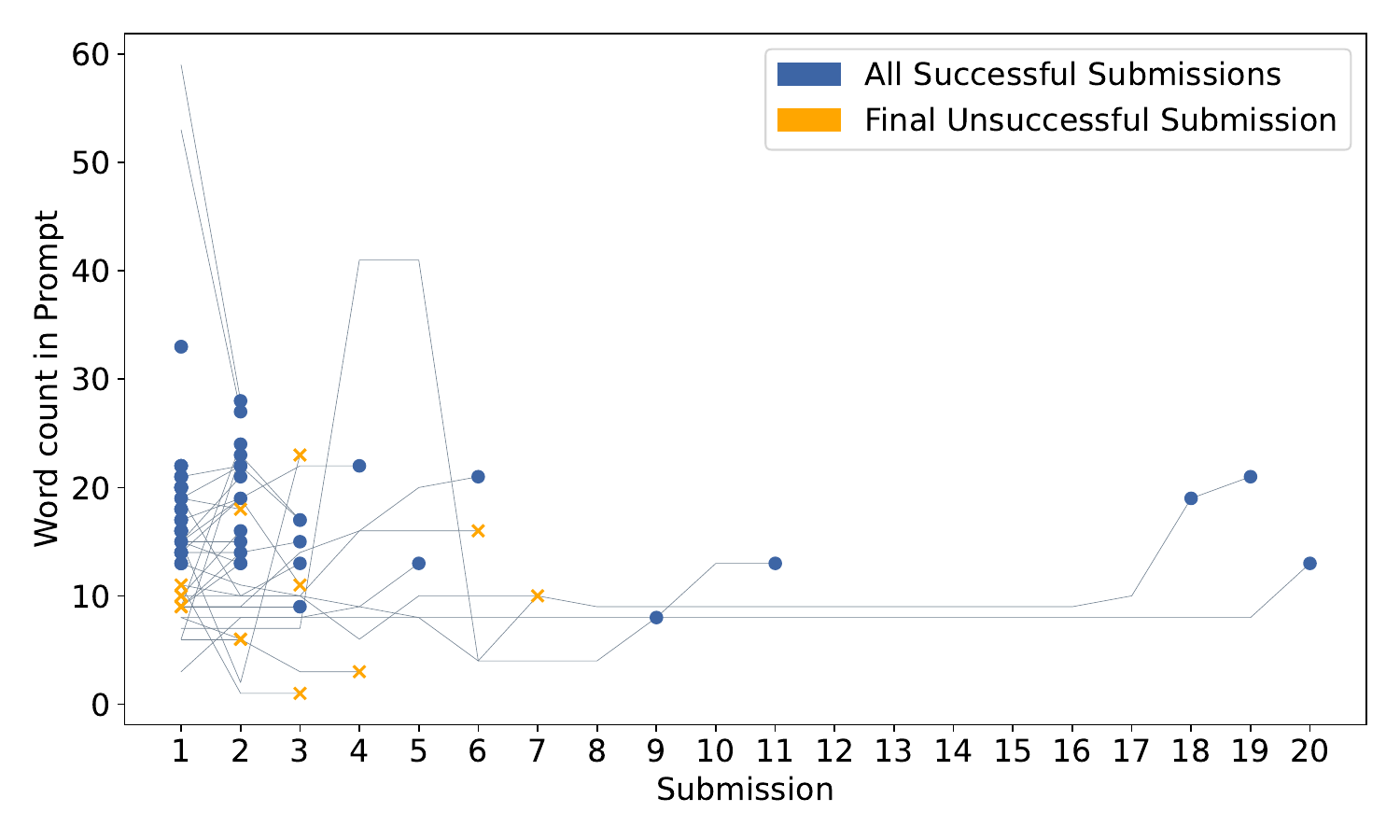}
  \caption{Each line represents all submissions made by a student for the CS1-1 problem.  Blue dots denote every successful submission; an orange X denote final unsuccessful submission.  Several students submit more than one successful prompt, indicating experimentation with the problem.}

  \label{fig:cs1-problem1}
\end{figure}
\begin{figure}
\centering
  \includegraphics[width=\linewidth]{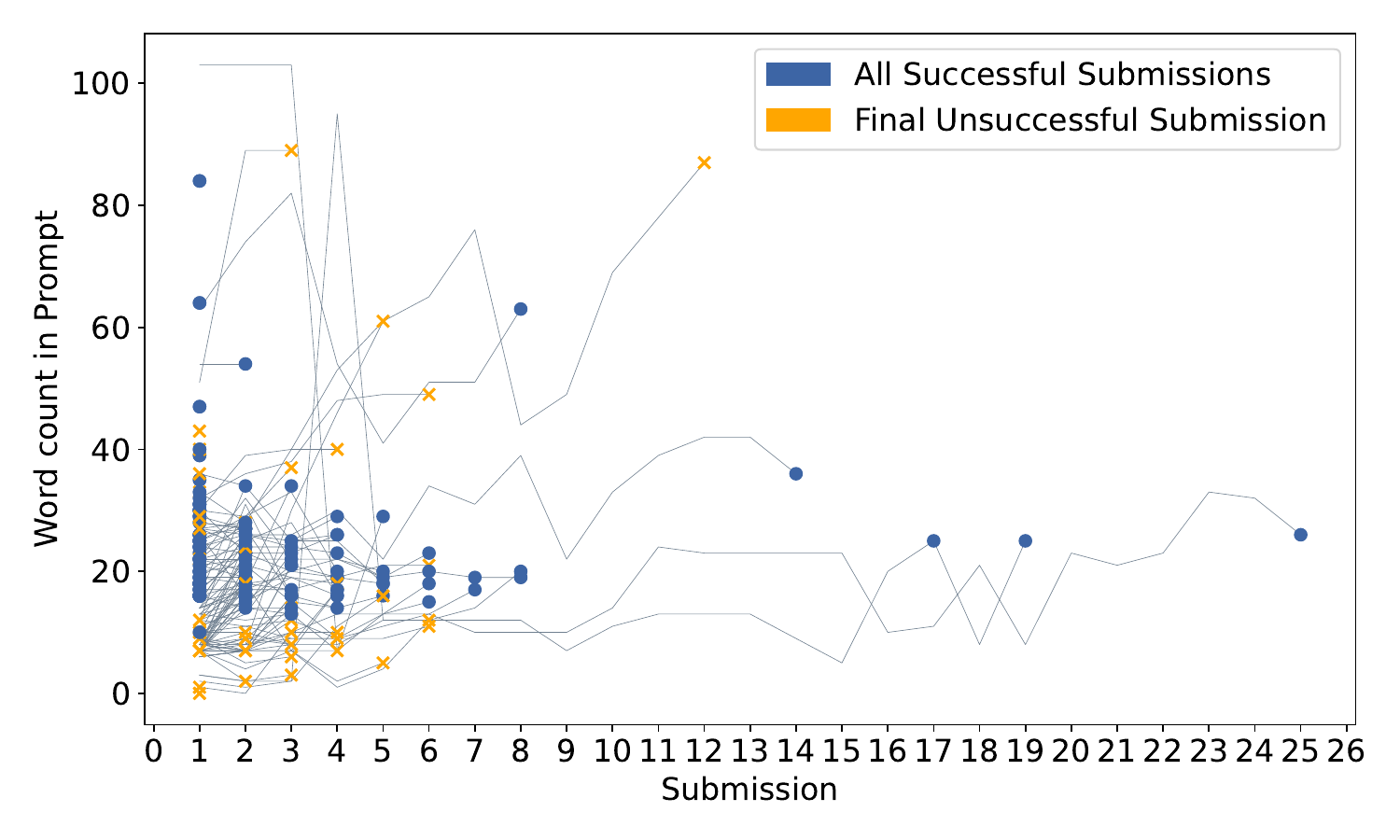}
  \caption{Each line represents all submissions made by a student for the CS2-1 problem. }
  \label{fig:cs2-problem1}
\end{figure}

Figures \ref{fig:cs1-problem1} and \ref{fig:cs2-problem1} show fine-grained submission patterns for the first problem in each course (CS1-1 and CS2-1, respectively).  Similar figures for all other problems are available as an online appendix\footnote{\url{https://osf.io/cw5gb/?view_only=343aeadc743047beb85764984ca1258b}}.  Each line on these figures represent the submissions made by one student, and illustrate how the word lengths of the prompts changed.  All successful submissions are highlighted with a blue dot; for students who did not solve the problem, the final unsuccessful submission is shown with an orange X.  Most students stopped working on a problem as soon as they solved it, although some continued working and experimenting with different prompts. 

In both figures, it is clear that many students solved the problem on their very first attempt (a single blue dot at submission 1).  An interesting observation here is the considerable variation in prompt length across these successful submissions.  It is likely that some of the longer prompts are not as succinct as they could be, which suggests some students may not be leveraging the power of the LLMs to the extent they could be.  As an example, the shortest successful prompts to CS2-2 and CS2-3 were the 12-word and 16-word prompts: \emph{``I want a function called initials which returns initials of the sentence''} and \emph{``Write me a Python3 function called repeat(list) which repeats the value according to its value''}.  In comparison, the longest successful prompts for these problems were 88 and 92 words, respectively.  Future variations of this activity could require that students submit working prompts that are less than some target length, to encourage them to be efficient with their word use.  Future work may also wish to reward students for the \emph{robustness} of their prompts, by calculating how frequently correct code is generated if the prompt is submitted multiple times.

\floatstyle{ruled}
\newfloat{Listing}{htbp}{lop}
\begin{Listing}
\caption{Three student-submitted prompts for CS1-3}
\label{listing:prompts_students}
\small
\raggedright
\textbf{Misinterpreting the problem:} \break
Write me a Python program that does the following:\break
1. Prompts the user to enter five decimal numbers (1dp) between 1.0 and 10.0 separated by spaces. \break
2. Chooses three of these numbers using the following rules: a number chosen be different from the previously chosen numbers and each subsequently chosen value must be within 0.5 of its predecessor. If the user has not provided numbers that sufficiently meet this criteria, call them an idiot and prompt them for another five values. \break
3. Find the average of these numbers and round the result to 2dp. Precede this result with the numbers chosen.

\vspace{-2mm}
\hfill \break
\textbf{Parroting the tests:} \break
A Python program requests the user "enter five decimal numbers (separated by spaces)".  In the first example the user inputs the five numbers 2.0 3.0 3.0 3.0 4.0 to which the program outputs 3.0.  In the second example the user inputs the five numbers 8.0 9.5 7.5 6.0 9.0 to which the program outputs 8.17 .  In the third example the user inputs the five numbers 4.0 6.5 8.0 7.0 6.0 to which the program outputs 6.5. Can you please replicate this program?

\vspace{-2mm}
\hfill \break
\textbf{Successful:} \break
Write me a Python program that takes five decimal number separated by spaces, and outputs the average of the 3 median numbers rounded to 2dp.

\end{Listing}

\subsection{Student Reflections on Prompt Problems}
\label{sec:reflections}
Of all the students who attempted at least one Prompt Problem in either course, a total of 153 chose to provide a response to the open-ended reflection question.
As this activity was new to students in both courses, we analyzed their feedback in combination. 
We report the main themes that emerged from our analysis below. 

\subsubsection{Exposure to new coding constructs}

Given that our evaluation was conducted early in both courses, the code that was generated would sometimes contain features that were unfamiliar to students.  For the most part, students commented very positively on this aspect, and a theme emerged around the way these problems would introduce students to new programming constructs and techniques.  As one CS1 student commented: \emph{``These exercises introduced me to new functions... so this method of writing code could help increase my programming vocabulary''}.  
Similar feedback was provided by students in the CS2 course, even though they had prior programming experience: \emph{``[Promptly] could find condensed ways to solve them using Python3's inbuilt functions, some even we have not been taught yet.''}


One student commented on the value of seeing both the structure and syntax of the code generated by the LLM: \emph{``The main benefit I gained ... was observing the logical structure of the programs that it created. 
In all three cases it used functions that I was previously unaware of, allowing me to gain an understanding of how they could be used and the correct syntax for implementing them.''}

\subsubsection{Enhancement of computational thinking}
To write prompts that are clear, it is often necessary to communicate problem solving steps, and this draws on computational thinking skills.  This is illustrated well by the following quote from a CS2 student:  \emph{``I do think that writing prompts for code is a good way of developing analytical and problem-solving thinking and skills as it forces you to think through the steps needed to take the input through to the output''}.  

Several participants found that writing prompts helped them improve their problem-solving skills, as they could focus on the logic required rather than low-level syntax: \emph{``I think while writing prompts for AI, we actually have to have a clear logic to break down the question and explain in plain words''} and \emph{``Gaining experience from writing prompts can help me become a more effective programmer by allowing me to generate the necessary code while focusing solely on the logic of the code I want to create''}.

\subsubsection{Resistance and negative feedback}
Although generally positive statements about the activity were far more common (e.g. \emph{``That was really fun! I loved the exercise and I feel like it would help me significantly in future labs''}), some students appeared resistant to taking part, citing fears about potential impacts on their creativity.  One student expressed: \emph{``I don't have much intention of using ChatGPT at the moment as I major in design and I have a strong belief in personal creativity''}.  Another was more blunt: \emph{``I refuse to use chatGPT for programming''}.  Over-reliance on AI generated outputs is a commonly cited concern within the education community, and several students commented on this aspect, including: \emph{``it is critical for students to learn the ability to write code independently rather than relying only on AI-generated answers''} and \emph{``I feel like it is too tempting of a tool to use through the labs and not learn and develop these skills yourself''}. These concerns align with previous work that has looked into students' opinions on AI code generation~\cite{prather2023its}. 

Further exploring these concerns is an essential avenue for ongoing work, given that some students appeared quite anxious about their future as computing professionals.  Upon reflecting on the Prompt Problems task, one student felt that there would no longer be a need for expertise in programming: \emph{``I don't think its a stretch to imagine that in the future `programmers' won't even be needed and will be replaced by someone who is able to write instructions for the program they want to make. I would be lying if I said I wasn't worried about the future of the majority of programming jobs.''}  Another student, in the CS2 course, commented on the emotional impact of the task and expressed rather bleak views of the future: \emph{``You have just ruined every piece of self esteem I had regarding coding. I know full well that it would have taken me around 35 minutes to figure out how to create those functions and that damn computer did it in seconds. Robots are going to own us within years.''} Overall, while most students reported finding Prompt Problems beneficial, particularly for exposure to new programming constructs and for strengthening computational thinking skills when explaining problems, a minority of students were both hesitant and concerned about the use of generative AI tools for learning programming.

%
%
\section{Discussion}

In contrast to other tools students use, such as compilers, learning to use LLMs presents unique challenges. 
For example, as educators we do not need to worry about teaching students that compilers might sometimes make a mistake, and yet the literature continues to document the difficulty students have with compiler error messages \cite{becker2019compilerWG, leinonen2023using}.  In contrast, identical input prompts to an LLM can produce different outputs, and these can sometimes be both syntactically and semantically incorrect.
Deliberate exposure to the inconsistencies of LLMs, such as through practice with Prompt Problems, can serve to highlight the importance of a ``critical eye'' in evaluating generated code and may help to moderate the potential for over-reliance on these tools.

Although our current tool evaluates prompt effectiveness in producing correct programs, it does not evaluate the efficiency of the prompts. Our unit tests consider only whether the given inputs are translated to the expected outputs.  A prompt could include irrelevant words and generate irrelevant code constructs, and as long as it still translates the given inputs to the expected outputs, our system will treat the task as completed successfully.  Future work should address how to go beyond effective prompts to efficient (and effective) prompts.

As this was our first experience deploying Prompt Problems to students, participation was optional.  In addition, students could only attempt a problem if they had successfully solved the previous problem.  As a result, there is likely considerable self-selection bias present in our data.  Nevertheless, early feedback from students was mostly very positive.  Future work should aim to expose Prompt Problems to a broader range of students, and provide incentives for their completion.

\subsection{Variations and Problem Design}
\label{discuss:variations}


There are various ways that Prompt Problems can be implemented, and our \textsc{Promptly} tool currently makes a number of trade-offs:  the problem must be solved by a single prompt and dialogue with the model is not allowed, it does not allow students to edit the code that is generated by the LLM, and it evaluates only a single response from the LLM at a time rather than generate and evaluate multiple responses.  We believe this provides a suitable experience for introductory level students, but many different variations are possible and should be explored -- including letting students engage in dialogue with the LLM and having the ability to edit the code that is generated.  Another particularly interesting variation of Prompt Problems is that instead of representing problems as inputs and outputs, as we have done, students could be presented with a code fragment and tasked with crafting a prompt that generates functionally equivalent code.  Such a variation combines aspects of code comprehension with prompt design. 

Finally, since prompt creation is a relatively new kind of task, it may be difficult for instructors to have an intuition for how difficult a particular Prompt Problem will be or when to utilize these types of problems.  By emphasizing problem solving over syntax, it may make it possible to introduce more complex problems sooner in a course.  Future work should explore more rigorously how best to integrate Prompt Problems alongside current teaching practices.

\section{Conclusion}

In this experience report we present a novel pedagogical approach, known as `Prompt Problems', designed to help students learn how to craft effective prompts for generating code using large language models (LLMs).  
We report our initial experiences deploying Prompt Problems to students for the first time using a novel tool we have developed, \textsc{Promptly}. 

We find that most students are able to solve Prompt Problems in just a few attempts, although some require 20 attempts or more, but when they do they construct a very wide variety of prompts.  For the most part, students report very positive experiences solving Prompt Problems, and value the exposure to new programming constructs and the enhancement of problem-solving skills.  However, a small number of students reported some hesitation about automated code generation, and a few even expressed anxiety about the future when seeing how powerful AI code-generating models can be.  Future work should investigate different variations of the approach we have described, and explore the right time to introduce students to the concept of prompt-based code generation.

\bibliographystyle{ACM-Reference-Format}
\bibliography{main.backup.long.bib}

\end{document}